\documentclass{nime-alternate}

\begin{document}
%
% --- Author Metadata here ---
\conferenceinfo{NIME'13,}{May 27 -- 30, 2013, KAIST, Daejeon, Korea.}

\title{Performing with a Mobile Computer System for Vibraphone}

\numberofauthors{1}
\author{
\alignauthor
Charles Martin\\
       \affaddr{Research School of Computer Science}\\
       \affaddr{Australian National University}\\
       \affaddr{Canberra, Australia}\\
       \email{cpm@charlesmartin.com.au}
}

\maketitle
\begin{abstract}
This paper describes the development of an \emph{Apple iPhone} based mobile computer system for vibraphone and its use in a series of the author's performance projects in 2011 and 2012.

This artistic research was motivated by a desire to develop an alternative to laptop computers for the author's existing percussion and computer performance practice. The aims were to develop a light, compact and flexible system using mobile devices that would allow computer music to infiltrate solo and ensemble performance situations where it is difficult to use a laptop computer.

The project began with a system that brought computer elements to \emph{Nordlig Vinter}, a suite of percussion duos, using an \emph{iPhone}, \emph{RjDj}, \emph{Pure Data} and a home-made pickup system. This process was documented with video recordings and analysed using ethnographic methods.

The mobile computer music setup proved to be elegant and convenient in performance situations with very little time and space to set up, as well as in performance classes and workshops. The simple mobile system encouraged experimentation and the platforms used enabled sharing with a wider audience.

\end{abstract}


\section{Introduction}
In 2010 I moved from Australia to Pite{\aa} in Northern Sweden to continue my studies in percussion and computer music. As part of this move I left most of my percussion and electronic music equipment behind in Australia. In this new life, I was extremely limited in the size and weight of equipment I could bring to concerts, lessons and even the shared practice studios, I was also working together with musicians with no experience with computer music.

I wanted to continue working on a previous project, \emph{Duet for Vibraphone and Computer}, which had used microphones and various reactive computer music elements on a laptop to allow the computer part to react to my playing on vibraphone. While this equipment had not been terribly heavy (in addition to the vibraphone, a laptop, audio interface, microphone, cable and microphone stand),  recent developments in mobile computer music inspired me use \emph{Apple's iPhone} and \emph{iPad} to create a system that might be able to fit inside my stick bag or backpack and integrate more elegantly with the vibraphone.

The concept of developing a simple, portable computer music system oriented for stage performance has previously been demonstrated by projects such as Audiopint~\cite{Merrill:2007la}, a rugged Linux computer system designed for live audio processing. Work by Tanaka~\cite{Tanaka:2010sp} and by Oh \emph{et al}~\cite{oh2010evolving} shows the musical possibilities of using  \emph{Apple's iOS} devices in musical performances and the accessibility of developing on this platform.

Another motivation of this work was to reduce the presence of a laptop computer in my percussion setups and increase my engagement with my main instrument, the vibraphone. A mobile phone based system could be made small enough to mount on the vibraphone itself just as Berdahl and Ju's Satellite CCRMA\cite{satelliteCCRMA2011} system could be built inside their controller instruments.

This project is part of a larger study that was the topic of my master's thesis \emph{Mobile Computer Music for Percussionists}, completed in June 2012 at Lule{\aa} University of Technology\cite{Martin:2012yq}. This thesis also includes a study of mobile computer instruments in an ensemble context.

\section{Research Questions}
The goal of this project was to address four research questions about mobile computer music:

\begin{enumerate}
\item \emph{Heaviness.} Can computer music setups be made more simple, elegant and convenient using mobile devices?

\item \emph{Shareability.} How can mobile computer music instruments be made accessible to a non-programmer percussion ensemble, and what creative processes can be used to explore them?

\item \emph{Playability.} How can the affordances of mobile music devices be used to create playable instruments for percussionists?

\item \emph{Performance practice.} What new performance practices are enabled or demanded when complementing acoustic percussion instruments with mobile music devices?
\end{enumerate}

\section{Research Method}

The research questions were adressed qualitatively through ethnographic\footnote{Ethnography is a qualitative research method for studying cultural phenomena. The researcher conducts fieldwork to collect notes, audio and video recordings, and images relating to the phenomena, an active and subjective method. ``The open-ended nature of the ethnographic approach is particularly suitable for active discovery and exploration''~\cite{Kruger:2008fk}. Analysis of the data can be an iterative process with multiple phases of field work and analysis to refine the research question. Conclusions are drawn inductively from the data gathered.} analysis of the process of developing a prototype mobile computer music system for vibraphone and using it in two performances projects: adding live computer elements to \emph{Nordlig Vinter} a suite of compositions for vibraphone and marimba, and \emph{Drums + Gadgets} an improvised performance developed in Columbus, Ohio with Noah Demland.

\subsection{Design Specifications}

The specifications for this system were for it to run a programmable computer music environment, have a microphone to input sounds from the vibraphone and have a stereo line out. However, in order to satisfy the motivations, the system needed to be small, light and elegant as well! In particular, the whole system needed to be small enough to sit on the end rail of a vibraphone without extra stands or tables. As borrowed instruments are frequently used for performances, the system could only use Velcro or Blu-Tack attachments to the instrument. The system needed to be compact enough to fit into a mallet bag (along with a few sets of mallets!) and should require the fewest possible cables so that the instrument can be relatively mobile on stage. Finally, the system needed to be cheap and simple enough that multiple versions could be made to share with an ensemble.

\section{Results}

\subsection{A Prototype Computer Music System for Vibraphone}

\begin{figure}[htbp]
   \centering
   \includegraphics[width=0.90\columnwidth]{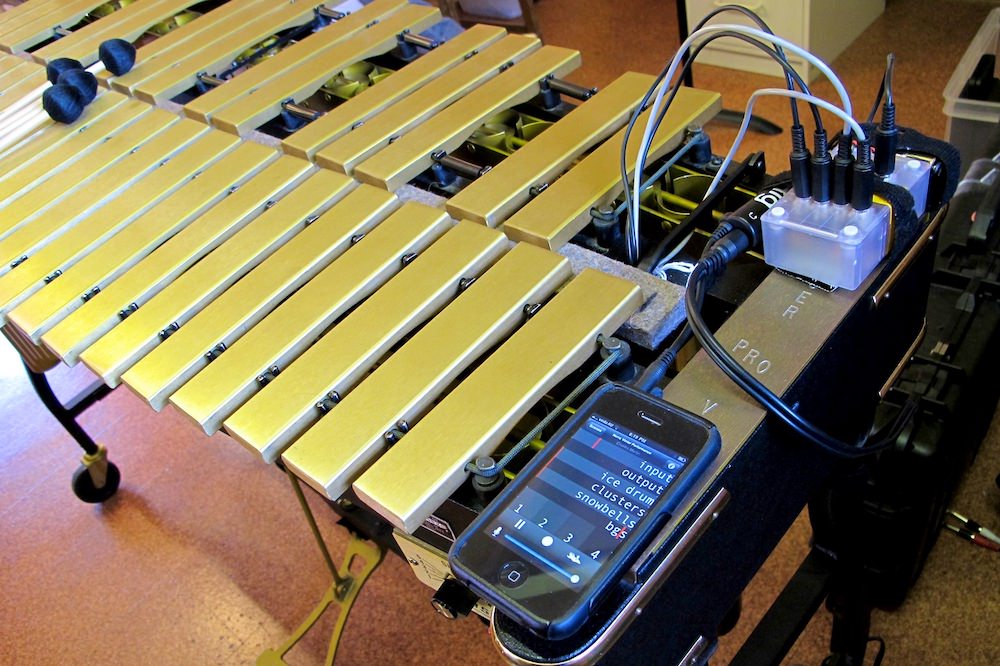}
   \caption{The prototype vibraphone/\emph{iPhone} system installed on the instrument.}
   \label{fig:vibraphoneinterface-installed}
\end{figure}

\begin{figure}[htbp]
   \centering \includegraphics[width=0.90\columnwidth]{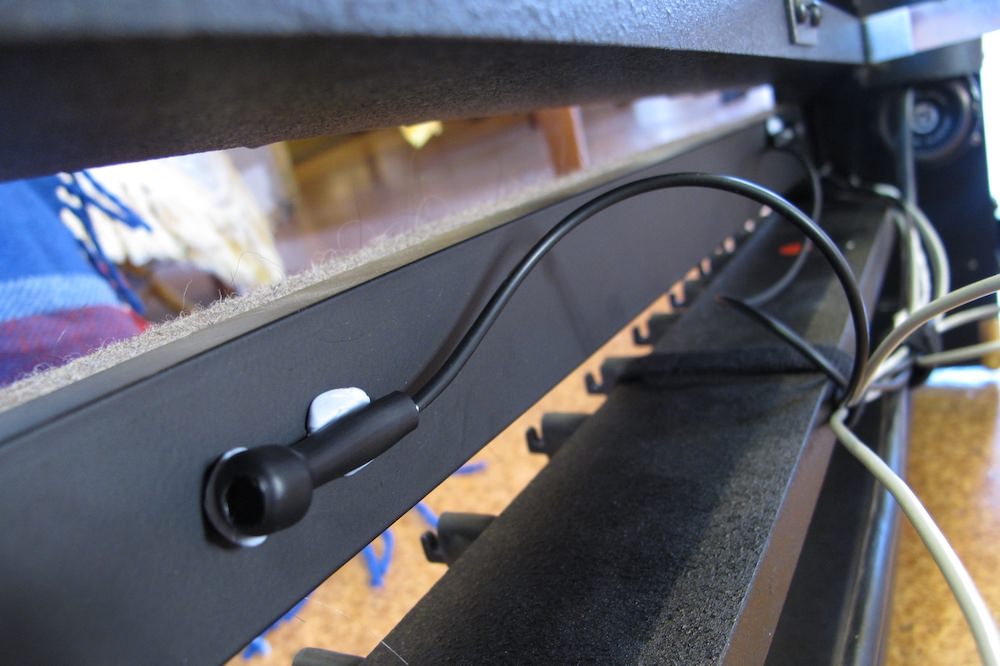}
   \caption{The microphones attached to the bottom of the damper bar.}
   \label{fig:vibraphoneinterface-microphones}
\end{figure}

To meet these specifications, the system was designed around an \emph{iPhone} running the \emph{RjDj}\footnote{\emph{RjDj} has now unfortunately been discontinued by the developer.}\cite{rjdjwebsite} app which had the ability to run custom \emph{Pure Data} patches called ``scenes''. The system contained the following elements:

\begin{enumerate}
\item Four electret microphones attached to the vibraphone's damper pedal with blu-tack (shown in detail in figure \ref{fig:vibraphoneinterface-microphones}).

\item A battery powered preamp and power supply, inspired by Collins' schematics~\cite{Collins:2009fk}, for the microphones that mixes them down to a mono signal. (shown in figure \ref{fig:vibraphoneinterface-microphones}).

\item An \emph{IK Multimedia iRig} dongle\footnote{\url{http://www.ikmultimedia.com/irig}} which separates the microphone input from the headphone outputs in the \emph{iPhone}'s mini jack connector.

\item An \emph{iPhone} running \emph{RjDj}.

\item A custom \emph{RjDj} scene containing the computer music elements for \emph{Nordlig Vinter} (programmed in \emph{Pd}).

\end{enumerate}

The system was completely battery powered and the phone, preamp and batteries were small enough to sit on the end rail of the vibraphone. The array of four electret microphones could be Blu-Tacked underneath the bars of the vibraphone hidden from the audiences view and out of the way of the performer. The only ``wired" aspect of the system was a pair of stereo jacks to connect the sound output to a PA system. The total cost, apart from the \emph{iPhone}, was around A\$100.

After this prototype was completed in September 2011, it was used in performances of \emph{Nordlig Vinter}, a suite of compositions for \emph{iOS} and percussion as well as in other collaborative improvisations and a classroom workshop.

\subsection{Nordlig Vinter: Works for Percussion and iOS}

\begin{figure}[htbp]
   \centering
   \includegraphics[width=5cm]{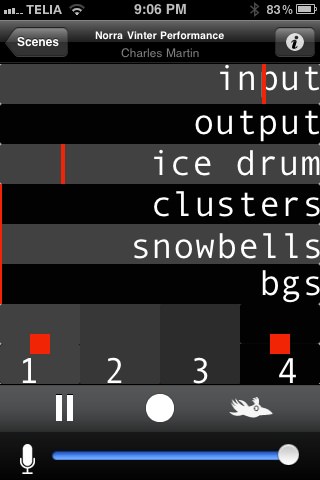}
   \caption{The \emph{RjDj} interface for performing \emph{Nordlig Vinter}.}
   \label{fig:norravinter-iphone-2}
\end{figure}

\emph{Nordlig Vinter} began in April 2011 as a collection of duo works for vibraphone and marimba inpired by the cold, dark, and snowy winter in Pite{\aa}. Some of the works were composed duos for marimba and vibraphone without a computer part and in others both players improvised over a background composition generated by an \emph{RjDj} scene (see figure \ref{fig:norravinter-iphone-2}). Two improvised pieces in the suite, \emph{Ice Drum} and \emph{Clusters}, were just for vibraphone and the computer music part. These two pieces also made use of the microphones in the vibraphone computer music system to apply effects to the vibraphone sound.

%The computer parts for \emph{Nordlig Vinter} were all generated by the \emph{RjDj} scene. These parts were generative, with the \emph{Pure Data} program at the core of the scene choosing where to play each note or sound based on an algorithm. They were also partly composed since timings for the entry of certain layers of sound and the overall timings of the works are hard-coded into the scene. 

The \emph{RjDj} scene was designed to require very little physical interaction during a performance. The scene had four buttons across the screen, the grey squares marked 1 to 4 in figure \ref{fig:norravinter-iphone-2}. The three compositions were started by the first three buttons. Red markers showed the current position of each of the compositions which stopped automatically after the markers reached the right-hand side of the screen. The fourth button toggled a reverb effect that could be applied to the vibraphone. The first two horizontal meters displayed the input and output volumes.

A recording of the \emph{Nordlig Vinter} suite was included with the original publication of this work in my master's thesis~\cite{Martin:2012yq}.

\subsection{Nordlig Vinter at Electrofringe}

\begin{figure}[htbp]
   \centering
   \includegraphics[width=0.90\columnwidth]{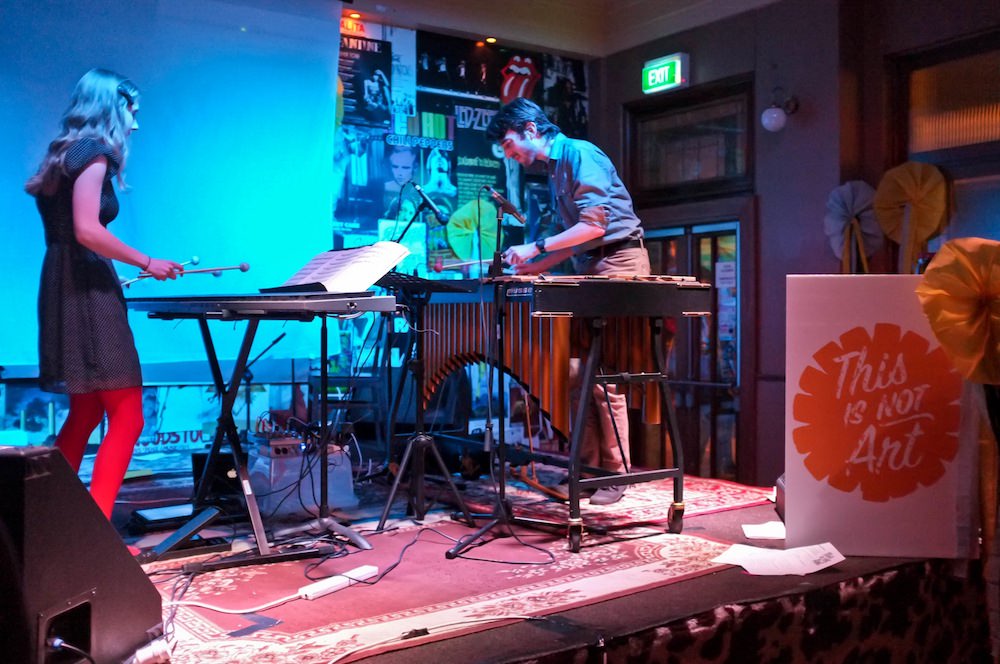}
   \caption{Christina Hopgood (left) and Charles Martin performing \emph{Nordlig Vinter, works for \emph{iOS} and percussion} at \emph{Electrofringe} 2011 in Newcastle, Australia. (Photo courtesy of Adam Thomas.)}
   \label{fig:norravinter-tina-2011}
\end{figure}

\emph{Electrofringe}\footnote{\url{http://electrofringe.net/}} is a yearly festival featuring experimental electronic arts in Newcastle, Australia. \emph{Nordlig Vinter} was performed as a percussion duo with Christina Hopgood in a performance night that included several other artists. The vibraphone computer music system was used in improvisations that connected each piece in the program.

This performance was a practical challenge for the system. The venue was a noisy and crowded pub with a ``rock and roll" style PA system and a small stage. There were a large number of performers for the evening with many different kinds of instruments and a variety of technical requirements.

Even in this challenging environment, we were able to set up the stage for our performance in around 5 minutes and performed after a very short sound check with only enough time to make sure all the instruments were working. The only technical difficulty was that due to feedback through the PA system's monitor speakers it was not possible to use the effects in the \emph{RjDj} scene. It's possible that with a more substantial sound check and more careful signal routing that this problem could have been eliminated.

At the same time as this concert, a version of the \emph{RjDj} scene for \emph{Nordlig Vinter} was publicly released with minor adjustments and a simplified interface that oriented it as a ``reactive" composition for a general audience. This scene could cycle through the generative compositions and process sounds through the \emph{iPhone}'s microphone and was available through the \emph{RjDj} app until October 2012.

\subsection{drums + gadgets}

\begin{figure}[htbp]
   \centering
   \includegraphics[width=0.90\columnwidth]{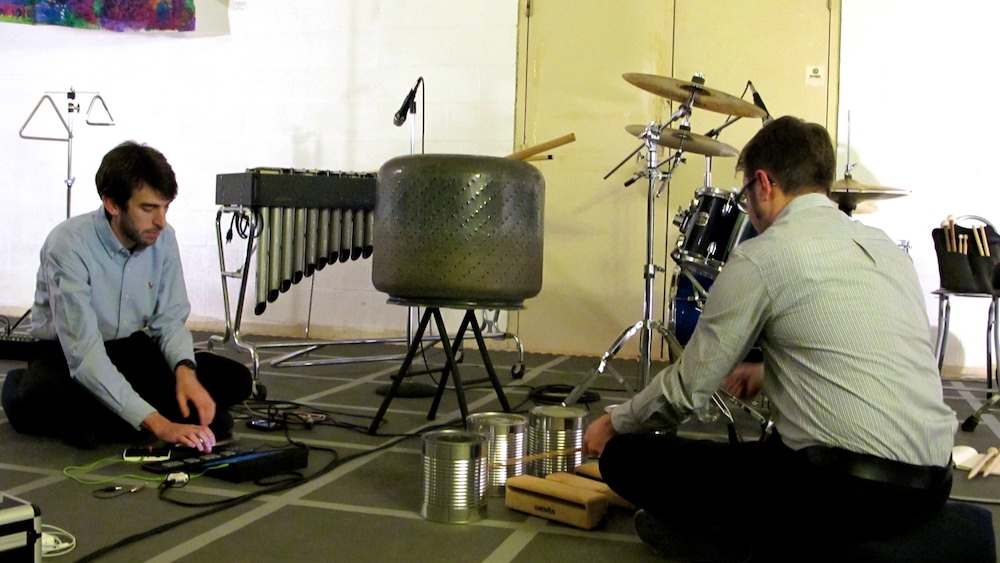} 
   \caption{Charles Martin and Noah Demland performing in \emph{drums + gadgets}, Columbus, Ohio, November 2011.}
   \label{fig:drumsgadgets-1}
\end{figure}

\emph{drums + gadgets} was an improvised concert of music for percussion and computer that I performed with Noah Demland in Columbus, Ohio in November 2011. Each of us contributed one of our semi-composed works which we arranged to play on vibraphone and drum set. We also created a number of improvised pieces using some of Noah's unique percussion instruments and my \emph{iOS} based computer instruments.

Taking elements from \emph{Nordlig Vinter}, Noah and I played a new interpretation of this work as an improvisation for vibraphone, drum set and \emph{iOS} using the vibraphone computer music system and the \emph{Nordlig Vinter} \emph{RjDj} scene\footnote{A video recording of this performance is available on Youtube: \url{http://youtu.be/fL4UxD_AwnQ}}. Rather than a series of composed pieces with improvisations and computer based elements in between, the work became a continuous improvisation interacting with the computer parts and recalling themes from the composition.

For the other improvised pieces, we agreed that I would play mainly \emph{iOS} based instruments and Noah would focus on percussion. In the resulting improvisations it turned out that some of my other \emph{RjDj} patches were useful for processing synthesiser programs on \emph{iPad} and that I could use a contact microphone with the \emph{iPhone} instead of the microphones that I had designed for the vibraphone. This contact microphone ended up being most useful on Noah's ``washing machine drum"\footnote{literally the inside of an old washing machine}. 

Some of these new developments were carried over into an improvisation workshop that I led for Noah's high school students. The students generally played guitars, keyboards and percussion in the workshops that focussed on composing and improvising ambient soundscapes, but I found that I was able to produce some stimulating background sounds for their explorations using the tools I had created in \emph{RjDj} and the electronic instruments used in the concert.
\vfill
%Overall the concert and workshop allowed me to perform with many of the \emph{iOS} computer music applications and electronic instruments I had been experimenting with. It was gratifying to experiment and develop some of these ideas as part of collaboration.

\section{Conclusions}

This project was successful overall with the mobile computer music system working well with the vibraphone in several performances.

\subsection{Heaviness}

The vibraphone computer music system was specifically designed to have minimal physical mass and stage presence.  The whole system could fit in a mallet bag, be set up on a vibraphone without any additional stands, and was completely battery powered. As demonstrated at \emph{Electrofringe}, the system could be set up in 5 minutes and work with minimal sound check although, as always, a more careful sound check will improve the musical result.

The use of \emph{RjDj} on an \emph{iPhone} as the only computer sound source resulted in a setup that is far simpler and quicker to setup than previous works using laptop computers running multiple computer music applications.

\subsection{Shareability}

One of the design considerations of the system was to share it with others in ensemble situations, this topic was investigated in a subsequent project to examine an ensemble performance using both perussion and computer instruments~\cite{Martin:2012fk}.

It is worth noting that the public version of \emph{Nordlig Vinter}'s \emph{RjDj} scene had been downloaded over 7000 times before \emph{RjDj} was discontinued by the developer. Although this is more a demonstration of the general popularity of \emph{iOS} devices than the attractiveness of the composition, it is remarkable how far the reach of mobile devices and app-stores are.

\subsection{Playability}

The design choice to use the \emph{iPhone} with \emph{RjDj} as the computing environment for this project put serious constraints on the visual interface. 
\emph{RjDj} included only very limited graphical user interface elements. 

As a result, the graphical interfaces for \emph{Nordlig Vinter} and other scenes developed for this project were very simple, displaying minimal information and  requiring the least possible interaction from the performer. In practice the \emph{iPhone}'s screen was so small and the information so limited that it was rarely the focus of my attention during performances. These limitations actually increased the playability of the whole system by allowing the player to focus on performance with minimal interruption or distraction. 

\subsection{Performance Practice}

The \emph{Nordlig Vinter} \emph{RjDj} scene created for the vibraphone computer music system and used in these two performances was intended to play background sounds for improvisations and for effects processing of the vibraphone sounds.

The concert at \emph{Electrofringe} featured through-composed pieces as well as  improvisations motivated by the sounds produced by the \emph{RjDj} scene. At \emph{drums + gadgets}, the lines between these performance modes became more blurred with background sounds and effects from the \emph{RjDj} scene continually used over the improvised performances. 

When creating \emph{drums + gadgets} with Noah Demland, we discovered that other instruments could be processed with effects in my \emph{RjDj} scenes, using direct inputs for electronic instruments, the vibraphone microphones or contact microphones. The simple modularity of this system enabled and encouraged this kind of experimentation and evolution in performance practice with a corresponding result in creativity.
%\vfill
%\columnbreak

\section{Further Developments}
Subsequent to this research, the \emph{Nordlig Vinter} scene was released as a native \emph{iOS} app\footnote{\url{https://charlesmartin.com.au/blog/2013/04/26/nordlig-vinter-app}} using the \emph{libpd}~\cite{Brinkmann:2011fy} library to continue updates and distribution after \emph{RjDj} was discontinued. The home-made preamp and headphone breakout in the system was replaced with a commercial product, \emph{IK Multimedia}'s \emph{iRig Pre}.

\section{Acknowledgments}
This research was supported by the Department of Arts, Communication and Education at Lule{\aa} University of Technology as part of my master's studies there, and was supervised by Stefan \"Ostersj\"o and Anders {\AA}strand.

\bibliographystyle{abbrv}
\bibliography{nime2011-references}
%%% Place this command where you want to balance the columns on the last page. 
%\balancecolumns 
\end{document}